# The structure of borders in a small world

C. Thiemann[*,†], F. Theis[‡], D. Grady[*], R. Brune[*,†] & D. Brockmann[*]

[*] *Northwestern University, Dept Eng Sci & Appl Math & Northwestern Institute on Complex Systems, Evanston, IL 60208 USA*
[†] *Max-Planck Institute for Dynamics and Self-Organisation, Bunsenstr. 10, 37073 Göttingen, Germany*
[‡] *Institute of Bioinformatics and Systems Biology, Helmholtz Zentrum München, German Research Center for Environmental Health, Neuherberg & Department of Mathematical Science, Technische Universität München, Garching, Germany*

**Geographic borders are not only essential for the effective functioning of government, the distribution of administrative responsibilities and the allocation of public resources[1,2], they also influence the interregional flow of information, cross-border trade operations[3], the diffusion of innovation and technology[4], and the spatial spread of infectious diseases[5-10]. However, as growing interactions and mobility across long distances[11], cultural, and political borders continue to amplify the small world effect[12,13] and effectively decrease the relative importance of local interactions, it is difficult to assess the location and structure of *effective* borders that may play the *most significant* role in mobility-driven processes. The paradigm of spatially coherent communities may no longer be a plausible one, and it is unclear what structures emerge from the interplay of interactions and activities across spatial scales[11,14-16]. Here we analyse a multi-scale proxy network for human mobility that incorporates travel across a few to a few thousand kilometres. We determine an effective system of geographically continuous borders implicitly encoded in multi-scale mobility patterns. We find that effective large scale boundaries define spatially coherent subdivisions and only partially coincide with administrative borders. We find that spatial coherence is partially lost if only long range traffic is taken into account and show that prevalent models[17-19] for multi-scale mobility networks cannot account for the observed patterns. These results will allow for new types of quantitative, comparative analyses of multi-scale interaction networks in general and may provide insight into a multitude of spatiotemporal phenomena generated by human activity.**

Modern human communication and mobility has undergone massive structural changes in the past few decades[1]. The emergence of large-scale social and communication networks and more affordable long distance travel generated highly complex connectivity patterns among individuals in large scale human populations. Although geographic proximity still dominates human activities, they can no longer be characterised by local interactions only. For example, on the US air transportation network, more than

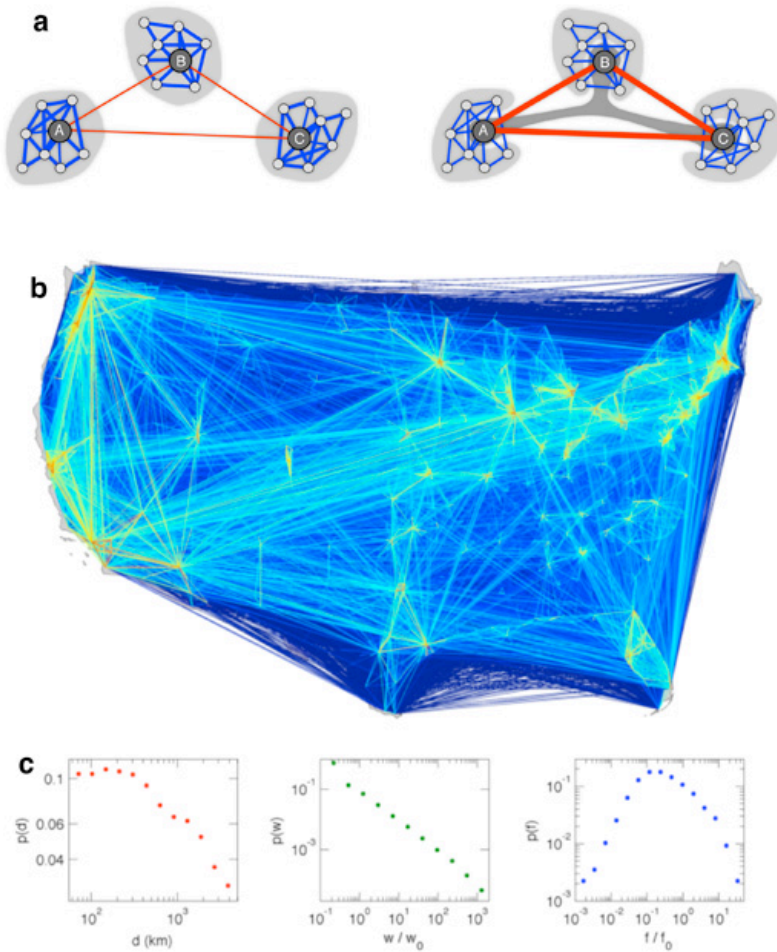

**Figure 1: Multi-scale human mobility. a**, An illustration of traffic flux between three large cities **A**, **B** and **C** and their respective local environments and plausible subdivisions into groups. **b**, A proxy network for multi-scale human mobility, illustrating the flux $w_{ij}$ of bank notes between 3,109 counties. Colour quantifies the intensity of flux, indicating high and significant mobility on short and long distances, respectively. **c**, Relative frequencies $p(d)$, $p(w)$, and $p(f)$ of distances $d_{ij}$, link weights $w_{ij}$ and vertex flux $f_i = \sum_j w_{ji}$ are distributed over several orders of magnitude, indicating the strong topological heterogeneity in the network. $w_0$ and $f_0$ are respectively the means of the $w_{ij}$ and $f_i$.

17 million passengers travel each week across long distances. However, including all means of transportation, 80% of all traffic occurs across distances less than 50 km[11,15]. The definition and quantification of meaningful geographic borders is particularly difficult in this intermediate regime of strong local and significant long distance interactions. Two plausible scenarios are illustrated in Fig. 1a, which shows the long range traffic flux between large cities A, B and C and local traffic within their respective geographic neighbourhoods. Taking traffic intensity as a measure of effective proximity, then depending on the ratio of local vs. long range traffic, two structurally different subdivisions could be meaningful. If short range traffic outweighs the long range traffic, local, spatially coherent subdivisions are meaningful. If long-range traffic dominates, subdividing into a single, spatially de-coherent urban community and three disconnected suburban modules is appropriate and effective geographic borders are difficult to define. Although previous studies identified community structures in long range mobility networks based on topological connectivity alone[20,21], this example illustrates that the precise interplay of mobility on all spatial scales as well as traffic intensity must be taken into account.

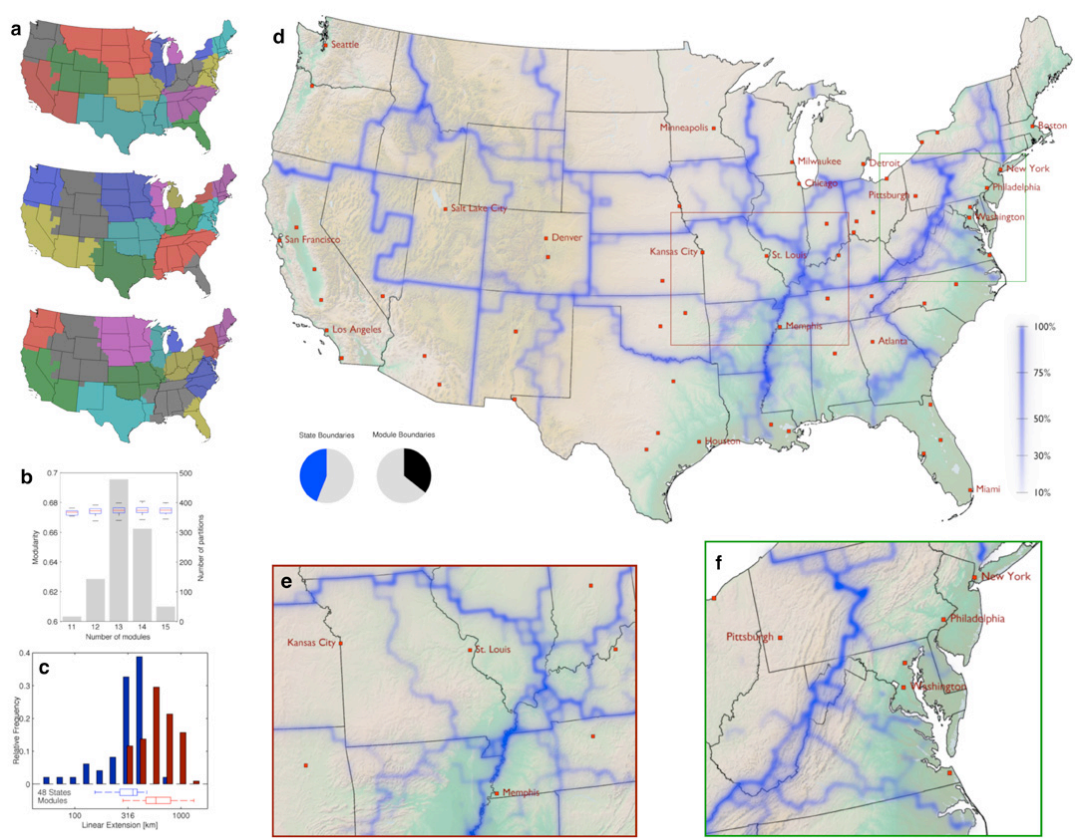

**Figure 2: Effective subdivisions and borders in the United States. a**, Subdivisions determined by maximising modularity $Q$ yielding (top to bottom) $Q = 0.6807, 0.6808$ and $0.6804$ and $k = 14$ (the same for all three). **b**, Distribution of $k$ and conditional distribution of $Q$ for 1000 subdivisions (ensemble mean $\overline{Q} = 0.674 \pm 0.0026$ standard deviations here and throughout). **c**, Distribution of linear extensions of states (mean $329 \pm 125$ km) and effective subdivision sizes ($644 \pm 215$ km). **d**, Emergence of effective borders by linear superposition of all maps in the ensemble (blue lines). Intensity encodes border significance (i.e. the fraction of maps that exhibit the border). Black lines indicate state borders. Although 44% of state borders coincide with effective borders (left pie chart), approximately 64% of effective borders do not coincide with state borders. **e**, Close-up on the Missouri region, showing the effective border between Kansas City and St. Louis that divides the state. **f**, Close-up on the Appalachian Mountains with corresponding border, which extends north to split Pennsylvania, where it is the most significant border in the map.

Obtaining comprehensive and complete datasets on human mobility covering many spatial scales is a difficult task. However, using the geographic movements of bank notes or cellphones is an excellent proxy of human mobility[11,16]. We construct a proxy network for human mobility from the movements of 8.97 million bank notes in the United States. Movement data was collected at the online bill-tracking system wheresgeorge.com. This dataset is more than an order of magnitude larger than one used in a previous study[11]. The network comprises 3,109 vertices representing the individual counties of the continental United States, and spans distances from 50 km (approximate average size of a county) to 3,000 km (linear extension of the US). The network is defined by the flux matrix $W$ whose elements $w_{ij} \geq 0$ (link weights) quantify the number of bills exchanged between counties $i$ and $j$ per unit time. $W$ is a symmetric weighted network. This strongly heterogeneous multi-scale mobility network is depicted in Fig. 1b showing that human mobility patterns are characterised by substantial short distance and significant long distance travel.

Based on the notion that two counties $i$ and $j$ are effectively proximal if $w_{ij}$ is large, we can use network-theoretic techniques[22] to identify a partition $P$ of the nodes into $k$ modules $M_n$ such that the intra-connectivity of the modules in the partition is high and inter-connectivity between them is low as compared to a random null model. A standard measure of the amount of community structure captured by a given partition $P$ is the modularity $Q(P)$ defined as[23,24]

$$Q = \sum_n \Delta F_n \qquad (1)$$

in which $\Delta F_n = F_n - F_n^0$ is the difference between $F_n$, the fraction of total mobility within the module $M_n$, and the expected fraction $F_n^0$ of a random network with an identical weight distribution $p(f)$. $Q$ cannot exceed unity; high values indicate that a partition successfully groups nodes into modules, whereas random partitions yield $Q \approx 0$. For large networks maximising $Q$ can only be performed approximately due to combinatorial explosion in the number of possible partitions. A variety of algorithms have been developed to systematically explore and sample the space of divisions in order to identify high-modularity partitions[21]. We apply an approximate, stochastic Monte-Carlo method (see also Supplementary Information) to find the optimal modular subdivision of the multi-scale mobility network by maximising the modularity. It has been pointed out that investigating ensembles of high-$Q$ subdivisions can convey better information of network structure, particularly for large networks where many different subdivisions exhibit similarly high values of $Q$. We therefore compute an ensemble of 1,000 divisions $P$ that all exhibit high modularity (i.e. $Q = 0.6744 \pm 0.0026$). Fig. 2a depicts the geographic representation of three sample partitions we find ($Q = 0.6807, 0.6806$ and $0.6804$) all comprising $k = 14$ modules. Note that, although modularity only takes into account the structure of the weight matrix $W$ and is explicitly blind to the geographic locations of nodes, the effective large-scale modules are spatially compact in every solution. Consequently, although long distance mobility plays an important role, the massive traffic along short distances generates spatial coherence of community patches of mean linear extension $l = 633 \pm 250$ km. Comparing these territorial subdivisions we note that although the modules in each map are spatially compact, possess almost identical modularity, and contain similar building blocks, they differ substantially in the set of modules they identify. It thus seems questionable to associate any quantitative meaning to individual patches and their borders; rather, this observation indicates that modularity may not be a useful order parameter for weighted, multi-scale mobility networks. However, a linear superposition of the set of maps exposes a complex structure of spatially continuous borders that represent a structural property of the entire ensemble of divisions (Fig. 2d). These borders are statistically significant topological features of the underlying multi-scale mobility network. Significance tests were performed using a null model generated by iterative local modifications of the border network (see Supplementary Information) and in general the border network significantly coincides with the state borders ($p < 0.001$). The superposition not only identifies the location of these borders but quantifies the frequency with which individual borders appear in the set of subdivisions. Note that these borders can effectively split states into independent patches, as with Pennsylvania, where the most significant border separates the state into Pittsburgh and Philadelphia spheres of influence. Other examples are Missouri, which is split into two halves, the eastern part dominated by St. Louis (also taking a pieces of Illinois) and the western by Kansas City, and the southern part of Georgia, which is effectively allocated to Florida. Also of note are the Appalachian mountains. Representing a real geological barrier to most means of transportation, this mountain range only partially coincides with state borders, but the effective mobility border is clearly correlated with it. Finally, note that effective patches are often centred around large metropolitan areas that represent hubs in the transportation network, for instance Atlanta, Minneapolis and Salt Lake City. We find that 44% of the administrative state borders are also effective boundaries, while 64% of all effective boundaries do not coincide with state borders (cf. Fig. 2d).

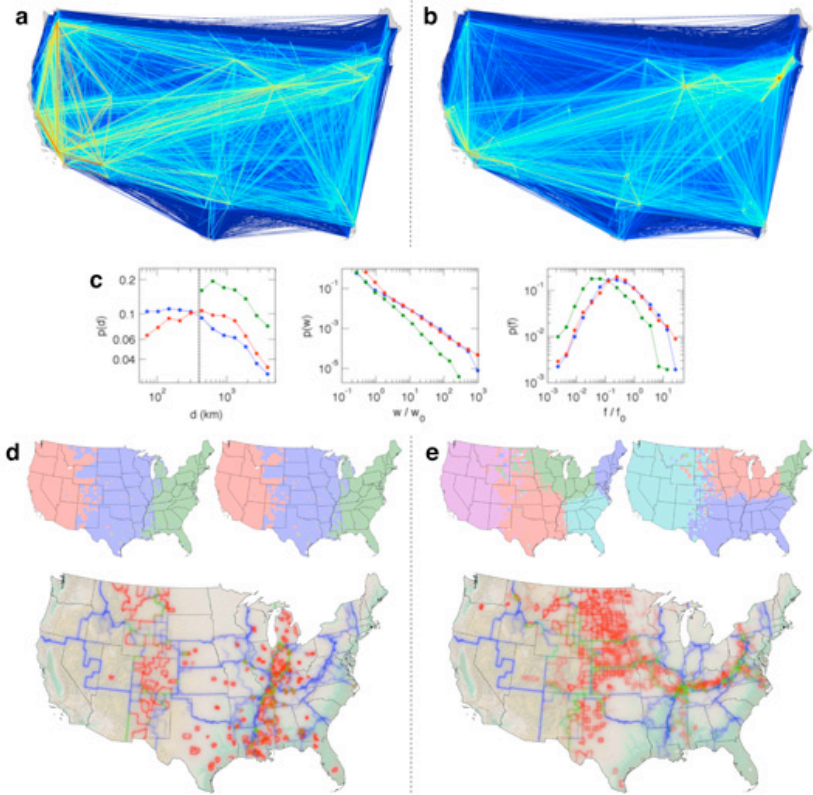

**Figure 3: Borders in long distance subnetwork and gravity model**. **a**, A subnetwork without short distance mobility ($d < 400$ km). 33.2% of the links were removed equivalent to 78.2% of the entire traffic. **b**, Gravity model network, generated according to Eq. (2). The parameters $\alpha = 1, \beta = 1, \mu = 0$ were chosen to maximise similarity with the original network. **c**, The functions $p(d)$, $p(w)$, and $p(f)$ for the long-range network in **a** (green), the gravity model network in **b** (red), and the original mobility network (Fig. 1b, blue). The dotted line indicates $d = 400$ km. **d**, Two typical partitions of the long-range network (top), and effective boundaries emerging from an ensemble of over 300 such maps (bottom). The effective borders of the reduced (red) and original (blue) network are compared and have only a minor overlap (green). **e**, Border structure for the gravity model network shown in the same way as **d**.

In order to test the degree to which short range connections dominate the structure of effective borders we generate an artificial network that lacks short range connections (Fig. 3a). Applying the same computational technique to locate and quantify effective spatial subdivisions, we find that removing short distance traffic has profound consequences for the spatial structure and coherence of divisions. We consistently find three independent modules that latitudinally split the US. As these three modules remain largely spatially coherent, we conclude that intermediate traffic inherits the role of short range mobility in generating spatial coherence. Although the removal of short links represents a substantial modification of the network, bootstrapping the original network randomly by the same amount (see Supplementary Information) has only little impact on the border structure depicted in Fig. 2d. We conclude that short distance mobility is a key factor in shaping effective borders.

We also investigate whether the observed pattern of borders can be accounted for by the prominent class of gravity models[17], frequently encountered in modelling spatial disease dynamics[19]. In these phenomenological models one assumes that the interaction strength between a collection of sub-populations with geographic positions $x_i$, sizes $N_i$, and distances $d_{ij} = |x_i - x_j|$ is given by

$$w_{ij} \propto N_i^\alpha N_j^\beta d_{ij}^{-(1+\mu)} \qquad (2)$$

in which $\alpha, \beta, \mu \geq 0$ are the parameters. Although their validity is still a matter of debate, gravity models are commonly used if no direct data on mobility is available. The key feature of a gravity model is that $w_{ij}$ is entirely determined by the spatial distribution of sub-populations. We therefore test whether the observed patterns of borders (Fig. 2d) are indeed determined by the existing multi-scale mobility network or rather by the underlying spatial distribution of the population. Fig. 3e illustrates the borders we find in a network that obeys equation (2). We generate this network such that the first order statistical similarity to the original networks is maximised, which sets the parameters $\alpha, \beta = 1$ and $\mu = 0$ (see Supplementary Information). Comparing this model network to the original multi-scale network we see that their qualitative properties are similar, with strong short range connections as well as prominent long range links. However, maximal modularity maps typically contain only five subdivisions with a mean modularity of only $\bar{Q} = 0.3543$. Because borders determined for the model system are strongly fluctuating (maps in Fig. 3e), they yield much less coherent large scale patches. However, some specific borders, e.g. the Appalachian rim, are correctly reproduced in the model. Because the model system produces significantly different patterns (see Supplementary Information for statistics), we conclude that the sharp definition of borders in the original multi-scale mobility network and the pronounced spatial coherence of the building blocks are an intrinsic feature of the real multi-scale mobility network.

In order to validate the structural stability of the observed patterns, we developed a new and efficient computational technique based on the concept of shortest path trees[25] (SPT) in combination with hierarchical clustering[26]. Like stochastic modularity maximisation, this technique identifies a structure of borders that encompass spatially coherent regions (Fig. 4b), but unlike modularity this structure is unique. The shortest path tree $T_i$ rooted at vertex $i$ is the union of all shortest paths originating at $i$ and ending at other vertices. SPTs are determined by the strength of connections in the network, and consequently are directly related to the dynamics that the network measures. The shortest path between two vertices is the path that minimises the effective distance $d = 1/w$ along the legs of the path. We quantify the common features of two trees by defining a measure of tree dissimilarity $z_{ij} = z_{SPT}(T_i, T_j)$ based on Hamming distance (see Supplementary Information). This measure indicates the amount of overlap taken in traveling from random locations in the network to the roots of the trees. Identical trees have a dissimilarity of zero and completely different trees have $z = N$, the number of nodes. In our data the $z$ values range from 2 to 240 (see Supplementary Information). We use a standard technique known as hierarchical clustering to compute a series of borders and consider a border more important if it appears earlier in the hierarchy. Unlike conventional clustering of the inverse weight matrix which requires adding some noise and produces a hierarchical structure that does not strongly correlate with the input, the set of borders computed by our method is an accurate representation of the underlying data (see Supplementary Information). In fact, although the method yields a unique sequence of topological segmentations, these geographic boundaries exhibit a strong correlation with those determined by modularity maximisation (Fig. 4a).

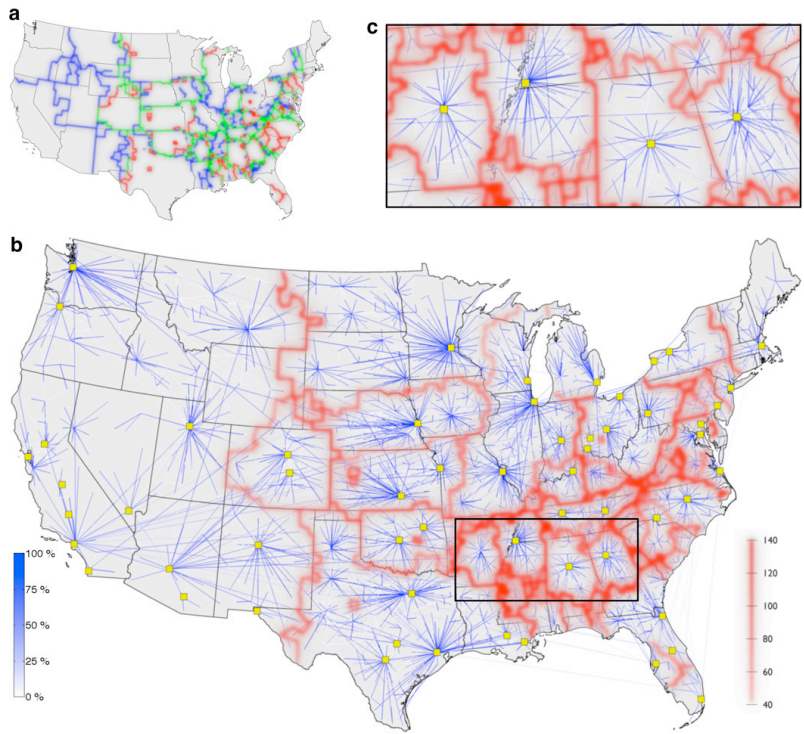

**Figure 4: Border identification from shortest-path tree clustering. a**, Comparison of borders from modularity maximisation (blue), SPT clustering (red) and their overlap (green). The cumulative topological overlap (see Supplementary Information) is 0.5282. **b**, Borders determined from SPT clustering. Intensity indicates height in the clustering tree (red colour bar). Links are drawn according to the frequency with which they appear in the set of all shortest-path trees (blue colour bar). **c**, Close-up of structure identified by SPT clustering. The marked cities are (left to right) Little Rock, AR; Memphis, TN; Birmingham, AL; and Atlanta, GA.

The key advantage of this method is that it can systematically extract properties of the network that match the observed borders. A way to demonstrate this is to measure the frequency at which individual links appear in the ensemble of all SPTs and to represent them as an effective network (Fig. 4b). Note that the prominence of links in this figure is not directly related to their traffic weight; instead, more prominent links are more likely to be used when traversing the network from one node to another by a hypothetical traveler. By virtue of the fact that the most frequently shared links between SPTs are local, short-range connections we see that the SPT boundaries enclose local neighbourhoods and that the boundaries fall along lines where SPTs do not share common features. Note that effective catchment areas around cities can be detected with greater precision than modularity (Fig. 4c), although some structures contain a single hub and some contain a 'dipole', and the west is detected as effectively a single community.

Based on our analysis we conclude that considerable geographic information is not only effectively encoded in human mobility networks, it can be quantified systematically using the techniques presented here. The quantitative identification, in particular, of geographic borders, and a comprehensive assessment of their significance will be very important for understanding dynamic processes driven by human mobility. We believe that because this general framework is suitable for a wide range of multi-scale human interaction networks for which the underlying effective borders are presently unknown, our results and the present framework may open the door for promising, quantitative, comparative investigations of numerous spatially distributed human activity patterns.